\documentclass[12pt]{article}
\usepackage{amsmath,amssymb,amsthm,latexsym,graphicx}
\newtheorem{theorem}{Theorem}
\newtheorem{lemma}[theorem]{Lemma}
\newtheorem{remark}[theorem]{Remark}
\newtheorem{corollary}[theorem]{Corollary}
\newtheorem{definition}[theorem]{Definition}
\newtheorem{proposition}[theorem]{Proposition}

\newcommand{\CC}{\mathbb{C}}
\newcommand{\TT}{\mathbb{T}}

\newcommand{\ZZ}{\mathbb{Z}}

\newcommand{\GG}{\Gamma}

\newcommand{\RR}{\mathbb{R}}

\begin{document}
\title{Quantum Graphs II: Some spectral properties of quantum
and combinatorial graphs}
\author{Peter Kuchment\\
Department of Mathematics\\
Texas A\& M University\\
College Station, TX, USA\\
e-mail: kuchment@math.tamu.edu}
\date{}
\maketitle

\begin{abstract}
The paper deals with some spectral properties of (mostly infinite)
quantum and combinatorial graphs. Quantum graphs have been
intensively studied lately due to their numerous applications to
mesoscopic physics, nanotechnology, optics, and other areas.

A Schnol type theorem is proven that allows one to detect that a
point $\lambda$ belongs to the spectrum when a generalized
eigenfunction with an subexponential growth integral estimate is
available. A theorem on spectral gap opening for ``decorated''
quantum graphs is established (its analog is known for the
combinatorial case). It is also shown that if a periodic
combinatorial or quantum graph has a point spectrum, it is
generated by compactly supported eigenfunctions (``scars'').
\end{abstract}
\section{Introduction}
We will use the name ``quantum graph'' for a graph considered as a
one-dimensional singular variety and equipped with a self-adjoint
differential ``Hamiltonian'', e.g.
\cite{ExS1,KoS1,Ku04_graphs1,QGraphs}. Such objects naturally
arise as simplified models in mathematics, physics, chemistry, and
engineering, in particular when one needs to consider wave
propagation through a ``mesoscopic'' quasi-one-dimensional system
that looks like a thin neighborhood of a graph. One can mention
among the variety of areas of applications of quantum graphs the
free-electron theory of conjugated molecules, quantum chaos,
mesoscopic physics (circuits of quantum wires), waveguide theory,
nanotechnology, dynamical systems, and photonic crystals. We will
not discuss any details of these origins of quantum graphs,
referring the reader instead to
\cite{ExS1,FK4,KoS1,KS,Ku01,Ku02,Ku04_graphs1,Ku04_spect,
KK99,KK02,N1,N2,N3,QGraphs} for further information, recent
surveys, and literature.

In this paper, which is a continuation of \cite{Ku04_graphs1}, we
present some results concerning spectra of quantum graphs, as well
as of their combinatorial counterparts. While the (combinatorial)
spectral graph theory has been around for quite some time
\cite{Chung,Col,CDS}, the spectral theory of quantum graphs has
not been developed well enough yet (see the collection
\cite{QGraphs} for recent developments and literature).

Let us describe the contents of the article. The next section
introduces the necessary notions concerning quantum graphs.
Section \ref{S:Schnol} contains a Schnol-Bloch type theorem. Such
theorems show how existence of a generalized eigenfunction with
some control on its growth (e.g., bounded) allows one to claim
that the corresponding point of the real axis is in fact in the
spectrum (or to estimate its distance from the spectrum). Section
\ref{S:gaps} deals with opening gaps in the spectrum of a quantum
graph by ``decorating'' the graph by an additional graph attached
to each vertex. Section \ref{S:bound} discusses point spectra of
periodic quantum graphs. It is shown that the corresponding
eigenspaces are generated by compactly supported eigenfunctions.
The results of all the sections have their counterparts in the
combinatorial setting as well.

It is interesting to note relations of the presented results with
their counterparts for PDEs. The Schnol type theorem is parallel
to the classical one known for PDEs \cite{Cycon,Glazman,Schnol},
except the integral formulation that we adopt, which extends its
applicability. The resonant gap opening procedure works to some
extent for PDEs as well \cite{Pav2}, but it is less clear and less
studied there. Finally, the discussion of the bound states for
periodic problems does not make much sense for PDEs, since
periodic second order elliptic operators with ``reasonable''
coefficients have absolutely continuous spectrum\footnote{Albeit
this statement is still not proven in complete generality yet, in
many cases it has been established (e.g., \cite{BirmSu,Fried2} and
references therein).}.

The reader should notice that although all the essential
ingredients of the proofs are presented, due to size limitations
the proofs are condensed and in some cases provided under some
additional restrictions that can be removed. A more detailed
exposition will appear elsewhere.

\section{Quantum graphs}\label{S:quantum_graphs}

A {\bf graph} $\Gamma$ consists of a finite or countably infinite
set of vertices $V=\{v_i\}$ and a set $E=\{e_j\}$ of edges
connecting the vertices\footnote{In this text we will be mostly
interested in infinite graphs}. Each edge $e$ can be identified
with a pair $(v_i,v_k)$ of vertices. Loops and multiple edges
between vertices are allowed. {\bf The degree (valence) $d_v$ of a
vertex $v$} is the number of edges containing the vertex and is
assumed to be finite and positive.
\begin{definition}\label{D:metr_graph} A graph $\Gamma$ is said to be
a {\bf metric graph}, if its each edge $e$ is assigned a positive
length $l_e \in (0,\infty )$ \footnote{Sometimes edges of infinite
length are allowed in quantum graphs. This is for instance the
case when one considers scattering problems.}.
\end{definition}
Each edge $e$ will be identified with the segment $[0,l_e]$ of the
real line, which introduces a coordinate $x_e$ along $e$. In most
cases we will denote the coordinate by $x$, omitting the
subscript. A metric graph $\Gamma$ can be  equipped with a natural
metric $\rho(x,y)$ and thus considered as a metric space. The
graph is not assumed to be embedded into an Euclidean space or a
more general Riemannian manifold. In some applications (e.g., in
modeling quantum wire circuits) such a natural embedding exists,
and then the coordinate $x$ is usually the arc length. In some
other cases (e.g., in quantum chaos), no embedding is assumed. All
graphs under the consideration are {\bf connected}.

We will also assume that the following additional condition is
satisfied:

$\bullet$ {\bf Condition A.} The lengths of all the edges are
bounded below and above by finite positive constants: $l_e \in
[l_0,L]$ for some $l_0>0, L<\infty$.

Condition A obviously matters for graphs with infinitely many
edges only. One can obtain some results without this condition as
well, but we will not address this issue here.

Now one imagines a metric graph $\Gamma$ as a one-dimensional
variety, with each edge equipped with a smooth structure, and with
singularities at the vertices:

\begin{figure}[ht]
\begin{center}
\scalebox{.3}{\includegraphics{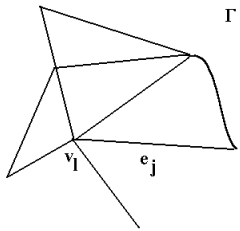}}
\end{center}
\caption{Graph $\Gamma$.}
\end{figure}

The reader should notice that the points of a metric graph are not
only its vertices, as it is normally assumed in the combinatorial
setting, but all intermediate points $x$ on the edges as well. So,
while a function on a combinatorial graph is defined on the set
$V$ of its vertices, functions $f(x)$ on a quantum graph $\GG$ are
defined along the edges (including the vertices). One can
naturally define the Lebesgue measure $dx$ on the graph.

We will sometimes assume that a {\bf root} vertex $o$ is singled
out (the results will not depend on the choice of the root). If
this is done, one can define a ``norm'' $\rho (x)$ of a point $x$
as
$$
\rho (x)=\rho(o,x).
$$
This allows us to define for any $r\geq 0$ the {\bf ball $B_r$ of
radius $r$}:
$$
B_r=\{x\in\Gamma\,|\, \rho (x) \leq r\}.
$$

The last step that is needed to finish the definition of a quantum
graph is to introduce a differential Hamiltonian on $\Gamma$. The
operators of interest in the simplest cases are the second arc
length derivative
\begin{equation}
f(x)\rightarrow -\frac{d^2f}{dx^2},  \label{E:deriv_2}
\end{equation}
or a more general Schr\"{o}dinger operator

\begin{equation}
f(x)\rightarrow \left( \frac 1i\frac d{dx}-A(x)\right)
^2f(x)+V(x)f(x). \label{E:electromagn}
\end{equation}
Here $x$ denotes the coordinate $x_e$ along each edge $e$.
\footnote{ Notice that in order to introduce the magnetic
operators, one needs to have graph's edges to be directed. This is
not required in the absence of the magnetic potential.}

Higher order differential and even pseudo-differential operators
arise as well (e.g., the survey \cite{Ku02} and references
therein). We, however, will concentrate here on second order
differential operators only.

In order for the definition of these self-adjoint Hamiltonians
operators to be complete, one needs to describe their domains. For
reasonable classes of potentials (e.g., measurable and bounded),
the natural conditions require that $f$ belongs to the Sobolev
space $H^2(e)$ on each edge $e$. One, however clearly also needs
to impose boundary value conditions at the vertices. These have
been studied and described completely using both the standard
extension theory of symmetric operators, as well as symplectic
geometry approach \cite{ExS3,Har,KoS1,Ku04_graphs1,N1,N2,N3}. The
simplest one is the so called Neumann condition\footnote{This name
seems to be more appropriate than the often used name Kirchhoff
condition, in particular for vertices of degree $1$ one obtains
the standard Neumann condition. Such conditions mean that the
quantum probability currents at each vertex add up to zero.}
\begin{equation}\label{E:Neumann_condit}
  \begin{cases}
    f & \mbox{ is continuous at each vertex } v \\
    \sum\limits_{v\in e} \frac{df}{dx_e}(v)=0 & \mbox{ at each vertex } v \
  \end{cases}
\end{equation}

\section{Schnol-Bloch theorems}\label{S:Schnol}

Schnol type theorems in PDEs (\cite{Schnol}, see also
\cite{Cycon,Glazman,Ku93,Shubin_bounded_geom}) treat the following
question. If there exists a non-zero $L_2$-solution of the
equation $Hu=\lambda u$, then clearly $\lambda$ is a point of the
point spectrum of $H$. Is there a similar test for detecting that
$\lambda$ belongs to the whole spectrum? Imagine that one has a
solution (a generalized eigenfunction) of a self-adjoint equation
$Hu=\lambda u$ and that one has some control of the growth of this
solution (e.g., it is bounded). When can one guarantee that
$\lambda$ is a point of the spectrum of $H$? For the
Schr\"{o}dinger equation in $\RR^n$ with a potential bounded from
below, the standard Schnol theorem \cite{Cycon,Glazman,Schnol}
says that existence of a sub-exponentially growing solution
implies that $\lambda \in \sigma (H)$. A version of this theorem
is known in solid state physics as the Bloch theorem
\cite{AM,Ku93,RS}: if $H$ is a periodic Schr\"{o}dinger operator,
then existence of a bounded eigenfunction corresponding to a point
$\lambda$ guarantees that $\lambda \in \sigma (H)$. On the other
hand, for the hyperbolic plane Laplace-Beltrami operator
$\Delta_H$, there is an infinite dimensional space of bounded
solutions of $\Delta_H u=0$. Indeed, using the Poincar\'{e} unit
disk model of the hyperbolic plane, one has
$\Delta_H=(1-|z|^2)^2\Delta$, where $\Delta$ is the Euclidean
Laplacian (e.g., Section 4 of the Introduction in \cite{Helgason},
or any other book on hyperbolic geometry). Thus, all bounded
harmonic functions $u$ on the unit disk (which form an
infinite-dimensional space) satisfy the equation $\Delta_H u=0$.
However, the point $0$ is still not in the spectrum of $\Delta_H$
(e.g., \cite{Lang}). This happens due to the exponential growth of
the volume of the hyperbolic ball of radius $r$. A similar Schnol
type theorem here would need to request some decay of the
generalized eigenfunction. The purpose of this section is
establishing a Schnol-Bloch type theorem for graphs.

Let $\Gamma$ be a rooted connected infinite quantum graph
satisfying the condition A and equipped with the Hamiltonian
$-\dfrac{d^2}{dx^2}$ and any self-adjoint vertex
conditions\footnote{More general Schr\"{o}dinger operators can be
treated similarly, see the remark after the theorem.}.

\begin{theorem}\label{T:Schnol} (A Schnol type theorem)
Let the graph $\Gamma$ satisfy the above conditions and $\lambda
\in \RR$. If there exists a function $\phi (x)$ on $\Gamma$ that
belongs to the Sobolev space $H^2$ on each edge, satisfies all
vertex conditions, the equation
\begin{equation} \label{E:Schnol}
-\frac{d^2\phi}{dx^2}=\lambda\phi \mbox{ for a.e. }x\in \Gamma,
\end{equation}
and the sub-exponential growth condition
\begin{equation}\label{E:subexp}
  \int\limits_{B_r}|\phi (x)|^2dx\leq C_\epsilon e^{\epsilon r}
\end{equation}
for any $\epsilon >0$, then $\lambda \in \sigma (H)$.
\end{theorem}

This theorem implies in particular the following

\begin{corollary}\label{C:Bloch} (A Bloch type theorem.)
Let the graph $\Gamma$ satisfy conditions of the Theorem and be of
a sub-exponential growth (i.e., the volume of $B_r$ grows
sub-exponentially). If there exists a bounded solution of the
equation (\ref{E:Schnol}), then $\lambda\in\sigma (H)$.
\end{corollary}

Simple examples show that existence of a bounded solution does not
guarantee that $\lambda\in\sigma (H)$, if the graph is of
exponential growth (i.e., a regular tree of degree 3 or higher).

{\em Proof of Theorem \ref{T:Schnol}}. Let us define for any $r>0$
the following compact subset $\Gamma_r$ of the graph: it consists
of all points of the edges with both ends in $B_r$. The following
inclusions hold:
\begin{equation}\label{E:gamma_r}
\Gamma_{r-L} \subset B_r \subset \Gamma_{r+L}.
\end{equation}
We hence conclude that the integral sub-exponential growth
condition (\ref{E:subexp}) holds if one replaces $B_r$ by
$\Gamma_r$. Let us introduce the function
\begin{equation}\label{E:growth function}
  J(r):=\int\limits_{\Gamma_r}|\phi (x)|^2dx.
\end{equation}
Given an $\epsilon >0$, one can find a sequence $r_k \to \infty$
such that
$$
J(r_k+L)\leq e^\epsilon J(r_k),
$$
otherwise one gets a contradiction with the sub-exponential growth
condition. We remind to the reader that each set $\Gamma_r$
consists of complete edges only.

Let $\theta(x)$ be any smooth function on $[0,l_0/4]$ such that it
is identically equal to $1$ in a neighborhood of $0$ and
identically equal to zero close to $l_0/4$. Here $l_0$ is the
lower bound for the lengths of all edges of $\Gamma$, which was
assumed to be strictly positive. We define a cut-off function
$\theta_k$ on $\Gamma$. It is equal to $1$ on $\Gamma_{r_k}$ and
to $0$ on all edges which do not have vertices in $\Gamma_{r_k}$.
We only need to define it along the edges that have exactly one
vertex in $\Gamma_{r_k}$. Let $e$ be an edge whose one vertex $v$
is contained in $\Gamma_{r_k}$. The function $\theta_k$ is defined
to be equal to $1$ along $e$ starting from $v$ till the middle of
the edge, then it is continued by an appropriately shifted copy of
$\theta (x)$ (which by construction will become zero at least at
the distance $l_e/4$ from the end of the edge), and stays zero
after that. Notice that due to the construction, any derivative of
the functions $\theta_k (x)$ is uniformly bounded with respect to
$k$ and $x \in \Gamma$. Besides, these functions are identically
equal to $1$ or $0$ around any vertex.

We can now construct a sequence of approximate eigenfunctions
$\phi_k (x)$ of the operator $H$ as follows:
$$
\phi_k (x)=\theta_k (x) \phi (x).
$$
One can notice that the functions $\phi_k (x)$ satisfy the same
boundary conditions that $\phi$ did, since the factors $\theta_k$
are identically equal to $1$ or $0$ around the vertices. This
implies that $\phi_k (x)$ belongs to the domain of $H$ in
$L^2(\Gamma)$. Besides, we clearly have
\begin{equation}\label{E:cut_growth_below}
\|\phi_k\|^2 \geq J(r_k).
\end{equation}
One also notices that the functions $\phi_k$ are supported in
$\GG_{r_k+L}$.

Let us now apply $H-\lambda$ to these test functions:
\begin{equation}\label{E:test}
(H-\lambda)\phi_k=\theta_k \left(-\phi^{\prime\prime}-\lambda \phi
\right) -2\theta_k^\prime \phi^\prime-\theta_k^{\prime\prime}
\phi=-2\theta_k^\prime \phi^\prime-\theta_k^{\prime\prime}.
\end{equation}
We have used here that $\phi$ satisfies (\ref{E:Schnol}).

Using the properties of the cut-off functions $\theta_k$, one gets
\begin{equation}\label{E:approx_eigen}
\|(H-\lambda)\phi_k\|^2  \leq  C \int\limits_{x\in \mbox{ supp
}\theta_k^\prime} \left(|\phi (x)|^2+|\phi^\prime (x)|^2 \right)dx
.
\end{equation}
Since the supports of the derivatives $\theta_k^\prime$ belong to
the interiors of the edges and are at a qualified distance from
the vertices, we have standard Schauder estimates for
$$
\int\limits_{x\in \mbox{ supp }\theta_k^\prime} |\phi^\prime
(x)|^2 dx
$$
by for instance the integral
$$
\int\limits_{\rho (x)\in [r_k+\frac{l_0}{4},r_k+L-\frac{l_0}{4}]}
|\phi (x)|^2dx.
$$
This leads to the estimate

\begin{equation}\label{E:estim_above_approx}
\begin{array}{c}
\|(H-\lambda)\phi_k\|^2 \leq C \int\limits_{\rho (x)\in
[r_k+\frac{l_0}{4},r_k+L-\frac{l_0}{4}]} |\phi (x)|^2dx\\
\leq C(J(r_k+L)-J(r_k)) \leq C (e^\epsilon -1)J(r_k) \leq  C
(e^\epsilon -1)\|\phi_k\|^2,
\end{array}
\end{equation}
where the constant $C$ does not depend on $k,\epsilon$. Since
$\epsilon >0$ was arbitrary, we conclude that $\lambda \in \sigma
(H) $. \qed

\begin{remark}\label{R:schnol}
\begin{enumerate}
\item If one has a generalized eigenfunction that satisfies (\ref{E:subexp})
for some fixed $\epsilon$, rather than arbitrary one as in the
theorem, one cannot conclude that $\lambda \in \sigma (H)$.
However, it is easy to modify the proof to estimate from above its
distance $\mbox{dist}(\lambda,\sigma (H))$ to the spectrum, which
when $\epsilon \to 0$ will reproduce the statement of the theorem.

\item The same result holds for more general Hamiltonian, e.g. for Schr\"{o}dinger
operators $-\dfrac{d^2}{dx^2}+q(x)$ with bounded from below
potentials $q(x)\geq q_0>-\infty$ and any self-adjoint vertex
conditions.

\item Analogous results, with essentially the same (a little bit simpler)
proofs hold for discrete operators on infinite combinatorial
graphs as well. One can notice then the relation of the Schnol
type theorems to the amenability properties of discrete groups and
graphs (e.g., the F\o lner condition) and to the notion of
infinite Ramanujan graphs.

The author will provide details concerning these remarks
elsewhere.
\end{enumerate}
\end{remark}

\section{Spectral gaps created by graph decorations}\label{S:gaps}
Existence of spectral gaps is known to be one of the spectral
features of high interest in the various fields ranging from solid
state physics to photonic crystal theory, to waveguides, to theory
of discrete groups and graphs. A standard way of trying to create
spectral gaps is to make a medium periodic (e.g.,
\cite{AM,Ku93,Ku01,RS}). This is why most of photonic crystal
structures that are being created are periodic. However,
periodicity neither guarantees existence of gaps (except in the
$1D$ case), nor it allows any easy control of gap locations or
sizes, nor it is a unique way to achieve spectral gaps. It has
been noticed by several researchers (the first such references
known to the author are \cite{Pav1,Pav2}), that spreading small
geometric scatterers throughout the medium (not necessarily in a
periodic fashion) might lead to spectral gaps as well. This has
been confirmed on quantum graph models in \cite{AEL,Exner}, and
finally made very clear and precise in the case of combinatorial
graphs in \cite{AiS}. It was proposed in \cite{AiS} that a simple
procedure of ``decorating'' a graph leads to a very much
controllable gap structure. We will show here that up to some
caveat, the same procedure works in the case of quantum graphs.
Let us describe the decoration procedure of \cite{AiS} adopted to
the quantum graph situation.

Let $\GG_0$ be a quantum graph satisfying the condition A and such
that the corresponding Hamiltonian is the negative second
derivative along the edges with the Neumann conditions
(\ref{E:Neumann_condit}) at the vertices\footnote{More general
conditions can also be considered.}. Let also $\GG_1$ be a {\bf
finite} connected quantum graph with the same type of the
Hamiltonian, with any self-adjoint vertex conditions. The graph
$\GG_1$ will be our ``decoration.'' We assume that a root vertex
$v_1$ is singled out in $\GG_1$. The decoration procedure works as
follows: The new graph $\GG$ is obtained by attaching a copy of
$\GG_1$ to each vertex $v$ of $\GG_0$ and identifying $v_1$ with
$v$ (see Fig. 2).
\begin{figure}[ht]
\begin{center}
\scalebox{.5}{\includegraphics{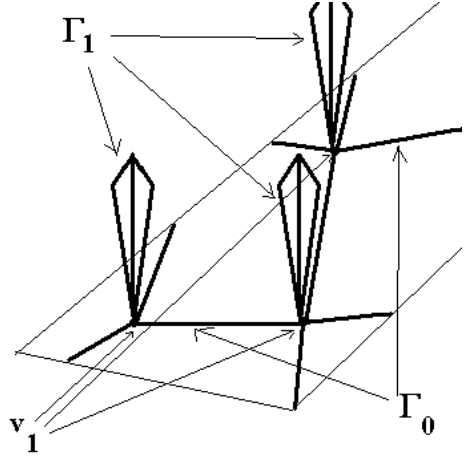}}
\end{center}
\caption{Decoration of a quantum graph $\GG_0$ by $\GG_1$.}
\end{figure}
Notice that there is a natural embedding $\GG_0 \subset \GG$. We
will denote by $V,V_0$, and $V_1$ the vertices sets of
$\GG,\GG_0$, and $\GG_1$ correspondingly. The Hamiltonian $H$ on
$\GG$ is defined as the negative second derivative on each edge,
with the Neumann conditions at each vertex of $\GG_0$ (including
the former $v_1$ vertices of the decorations) and the initially
assumed conditions on $V_1\backslash v_1$, repeated on each
attached copy of the decoration.

Dirichlet eigenvalues of each edge (which are clearly directly
related to the edge lengths spectrum) often play an exceptional
role in quantum graph considerations (see the discussions below).
Let $\{l_j\}$ be the lengths of the edges of the original graph
$\GG_0$. Then we define the {\bf Dirichlet spectrum $\sigma_D$ of
$\GG_0$} as the {\bf closure} of the set
$$ \mathop{\cup}\limits_{n\in
(\ZZ\backslash 0),j}\{\pi^2n^2/l_j^2\} \subset \RR.
$$
If the graph
$\GG_0$ is finite, no closure is required.

Let us also define the operator $H_1$ on the decoration graph
$\GG_1$ that acts as the negative second derivative on each edge
and satisfies the self-adjoint conditions assumed before on
$V_1\backslash v_1$ and zero Dirichlet condition at $v_1$.

We can now state the result of this section, which was previously
announced in \cite{Ku_isaac,Ku04_spect}. The conditions of the
theorem can be weakened, but we consider for brevity the simplest
case here, which seems already rather useful.

\begin{theorem}\label{T:decoration}
Let $\lambda_0 \in \RR \backslash \sigma_D$ be a simple eigenvalue
of $H_1$ with the eigenfunction $\psi$ such that the sum of the
derivatives of $\psi$ at $v_1$ along all outgoing edges is not
zero. Then there is a punctured neighborhood of $\lambda_0$ that
does not belong to the spectrum $\sigma(H)$ of $\GG$.
\end{theorem}
\proof We will prove here the theorem for the case of a finite
graph $\GG_0$ only. The case of an infinite graph is a little bit
more technical and will be considered elsewhere. The proof
consists of removing the decorations and replacing them by altered
vertex conditions. This is done simultaneously and the same way at
each vertex $v\in V_0 \subset V$, so we will describe it for one
vertex $v$, which will be identified with $v_1 \in \GG_1$.

Let us define a function that we will call Dirichlet-to-Neumann
function $\Lambda (\lambda)$ for $\GG_1$\footnote{This is in fact
the Dirichlet-to-Neumann map for $\GG_1$, if $v_1$ is considered
as this graph's boundary.}. It is defined in a punctured
neighborhood of $\lambda_0$ not intersecting $\sigma_D$ as
follows. If $\lambda \neq \lambda_0$ is a regular point of $H_1$,
one can uniquely solve the problem
\begin{equation}\label{E:DtN_equation}
{\begin{cases}
  -u^{\prime \prime}=\lambda u  \mbox{ on each edge of }\GG_1 \\
  u \mbox{ satisfies the prescribed boundary conditions on }
  V_1\backslash v_1\\
  u(v_1)=1
\end{cases}}
\end{equation}
We denote by $\Lambda (\lambda)$ the sum of the outgoing
derivatives of the solution $u(x)$ at the vertex $v_1$.
\begin{lemma}\label{L:DtN_function}
Under the conditions of the Theorem, the Dirichlet-to-Neumann
function $\Lambda (\lambda)$ is analytic  in a punctured
neighborhood of $\lambda_0$, with a first order pole (with
non-zero residue) at $\lambda_0$.
\end{lemma}
{\em Proof of the lemma.} Let $\psi$ be the eigenfunction of $H_1$
assumed in the statement of the theorem. We denote by $\Psi\neq 0$
the sum of outgoing derivatives of $\psi$ at $v_1$. Let also $f$
be a function on $\GG_1$ defined as follows: it is supported in a
small neighborhood of the vertex $v_1$ (so small that it does not
contain other vertices of $\GG_1$), is equal to $1$ near $v_1$,
and is smooth inside the edges. We also denote by
$R_{H_1}(\lambda)=(H_1-\lambda)^{-1}$ the resolvent of $H_1$. Then
we can represent the solution $u$ of (\ref{E:DtN_equation}) as
$\tilde{u}+f$, where
$$
\begin{array}{c}
\tilde{u}=-R_{H_1}(\lambda)(-f^{\prime\prime}-\lambda f)\\
=-(\lambda-\lambda_0)^{-1}<-f^{\prime\prime}-\lambda f,\psi>_{L_2(\GG)}\psi(x)+A(\lambda)\\
=-(\lambda-\lambda_0)^{-1}\Psi\psi(x)+A(\lambda).\\
\end{array}
$$
Here $A(\lambda)$ is analytic in a neighborhood of $\lambda_0$.
Noticing that the sums of the outgoing derivatives at $v_1$ of
both functions $u$ and $\tilde{u}$ on $\GG_1$ are the same, we see
that $\Lambda(\lambda)$ has a first order pole at $\lambda_0$ with
a non-zero residue. This proves the lemma.

Let now $\lambda_0$ be as in the theorem. Suppose that $u(x)$ is
an eigenfunction of $H$ corresponding to an eigenvalue $\lambda$
close to $\lambda_0$. For any vertex $v\in V_0$, we can solve the
equation $Hu=\lambda u$ on the decoration attached to $v$, using
$u(v)$ as the Dirichlet data. Then the sum of outgoing derivatives
of $u$ at $v$ along the edges of the decoration is equal to
$\Lambda(\lambda)u(v)$. Hence, the eigenfunction equation for $u$
on $\GG$ can be re-written on $\GG_0$ solely as follows:

\begin{equation}\label{E:reduction}
\begin{cases}
  -u^{\prime \prime}=\lambda u \mbox{ on each edge of }\GG_0 \\
  u \mbox{ is continuous at all vertices }v\in V_0 \\
  \sum\limits_{v\in e} \frac{du}{dx_e}(v) =
  -\Lambda(\lambda)u(v).
\end{cases}
\end{equation}

We will show now that (\ref{E:reduction}) is impossible for a
non-zero function $u$, if $\lambda$ is close to $\lambda_0$.
Indeed, with $\lambda$ being at a positive distance from the
Dirichlet spectrum $\sigma_D$ of all edges, standard estimates
give

\begin{equation}\label{E:resolvent estimate}
  \sum\limits_{e \in \GG_0} \|u\|^2_{H^2(e)} \leq C
  \sum\limits_{v \in V_0} |u(v)|^2.
\end{equation}

Now Sobolev trace theorem implies

\begin{equation}\label{E:trace}
  \sum\limits_{\{e\in \GG_0, v \in V_0|\,v\in e\}} |\frac{du}{dx_e}(v)|^2 \leq C
  \sum\limits_{v \in V_0} |u(v)|^2.
\end{equation}

Since $\Lambda(\lambda)$ has a pole at $\lambda_0$, for $\lambda$
and $\lambda_0$ sufficiently close, we get contradiction between
(\ref{E:trace}) and the last equality of (\ref{E:reduction}).\qed

\begin{remark}\label{R:gaps}
\begin{enumerate}
  \item As it was mentioned above, the proofs for the infinite
  case will be provided elsewhere.
  \item The proof shows that the decorations
  attached to each vertex do not have to be the same in order to achieve
  spectral gaps. One only needs to guarantee a uniform blow-up of all the
  Dirichlet-to-Neumann functions at each vertex when $\lambda \to
  \lambda_0$. One can also provide some estimates of the size of
  the gap.
  \item This theorem claims that spectral gaps are guaranteed to arise
around the spectrum of the decoration (with the Dirichlet
condition at the attachment point $v_1$), unless one deals with
the Dirichlet spectrum of $\GG_0$. Simple examples show that on
the Dirichlet spectrum one cannot guarantee a gap. For instance,
if $\GG_0$ contains a cycle consisting of edges of equal (or
commensurate) lengths, then the decoration procedure cannot remove
the eigenvalues that correspond to the sinusoidal waves running
around this loop (see Fig. 4). However, a modification of the
decoration procedure works even in the presence of Dirichlet
spectrum. One just needs to introduce some ``fake'' vertices along
the edges at appropriate locations and attach the decorations at
these new vertices as well. This will be described in detail
elsewhere.
  \item One can create gaps by a different decoration procedure
  rather than the one of \cite{AiS} described above. Namely, instead
  of attaching sideways the little ``flowers'' (or ``kites,'' as they
  were called in \cite{AiS}) as in Fig. 2, one could incorporate an internal
  structure into each vertex, putting a little ``spider'' there as shown in
  Fig. 3 below.
\begin{figure}[ht]
\begin{center}
\scalebox{.5}{\includegraphics{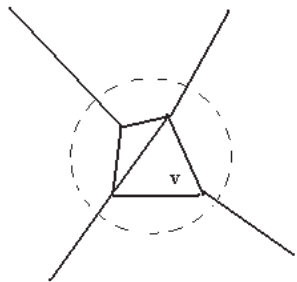}}
\end{center}
\caption{A ``spider'' decoration.}
\end{figure}
This graph decoration procedure was probably used explicitly for
the first time in \cite{AEL} (see also \cite{Exner}) for the same
purpose of gap creation. It will be shown elsewhere how gaps can
be created using this construction (the Dirichlet spectrum plays a
distinguished role there as well).
\end{enumerate}
\end{remark}

\section{Bound states on periodic graphs}\label{S:bound}

It is ``well known'' (albeit still not proven for the most general
case) that elliptic periodic second order operators in $\RR^n$
have no point spectrum\footnote{This is not true for higher order
operators \cite{Ku93}.}. In fact, their spectra are absolutely
continuous. In the case of Schr\"{o}dinger operators with periodic
electric potentials, this constituted the celebrated Thomas'
theorem \cite{Thomas} (see also \cite{Ku93,RS}). There has been a
significant progress in the last decade towards proving this for
the general case. One can find the description of the status of
this statement for the general elliptic periodic operators in
\cite{BirmSu,Fried2,Ku01,KuL}. The validity of this theorem is
intimately related to the uniqueness of continuation property
(that is why it fails for higher order operators), which does not
hold on graphs. It is well known that bound states, and even
compactly supported eigenfunctions can easily be found in
combinatorial and quantum graphs, whether periodic or not. If, for
instance the quantum graph has a cycle with commensurate lengths
of the edges, one can easily create a sinusoidal wave supported on
this loop only (see Fig. 4).
\begin{figure}[ht]
\begin{center}
\scalebox{.3}{\includegraphics{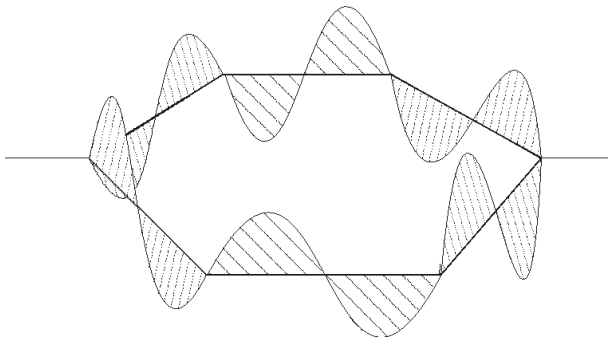}}
\end{center}
\caption{A loop bound state.}
\end{figure}
The question arises whether any other causes exist besides
compactly supported eigenfunctions, for appearance of the pure
point spectrum on periodic graphs. It has been shown previously by
the author \cite{Ku89} that in the case of combinatorial periodic
graphs, existence of bound states implies existence of the
compactly supported ones. In fact, the eigenfunctions with compact
support generate the whole eigenspace. We will show here that the
same holds true for periodic quantum graphs as well.

One should note that point spectrum can arise for different
reasons on graphs that are not periodic, e.g. on trees. For
instance, one can have bound states on infinite trees with
sufficiently fast growing branching number \cite{Solomyak}.

We will consider an infinite combinatorial or quantum graph $\GG$
with a faithful co-compact action of the free abelian group
$G=\ZZ^n$ (i.e., the space of $G$-orbits is a compact graph).

Let us treat the combinatorial case first, so let $\GG$ be a
combinatorial graph and $A$ a $G$-periodic finite difference (not
necessarily self-adjoint) operator of a finite order acting on
$l_2(V)$. Here, as before, $V$ is the set of vertices of $\GG$.
The first half of the following result is proven in \cite{Ku89}:
\begin{theorem}\label{T:bound_combinat}
If the equation $Au=0$ has a non-zero $l_2(V)$ solution, then it
has a non-zero compactly supported solution. Moreover, the
compactly supported solutions form a complete set in the space of
all $l_2$-solutions.
\end{theorem}
Since this formulation is more complete than the one in
\cite{Ku89}, we provide its brief proof here.

\proof We will need to use the basic transform of Floquet theory
(e.g., \cite{Ku93,RS}). Namely, for any compactly supported (or
sufficiently fast decaying) function $u(v)$ on $V$, we define its
Floquet transform
\begin{equation}\label{E:Floquet}
  u(v)\mapsto \hat{u}(v,z)=\sum\limits_{g\in\ZZ^n} u(gv)z^g,
\end{equation}
where $gv$ denotes the action of $g\in\ZZ^n$ on the point $v\in
V$, $z=(z_1,...,z_n)\in (\CC\backslash 0)^n$, and
$z^g=z_1^{g_1}\times...\times z_n^{g_n}$. We will also denote
$\hat{u}(v,z)$ by $\hat{u}(z)$, where the latter expression is a
function on $W$ depending on the parameter $z$. Here $W$ is a
(finite) fundamental domain of the action of the group $G=\ZZ^n$
on $V$. Notice that images of the compactly supported functions
are exactly all finite Laurent series in $z$ with coefficients in
$\CC^{|W|}$,

We will also need the unit torus
$$
\TT^n=\{z\in\CC^n\,|\,|z_j|=1,j=1,...,n\}\subset\CC^n.
$$

It is well known and easy to establish \cite{Ku89,Ku93,RS} that
the transform (\ref{E:Floquet}) extends to an isometry (up to a
possible constant normalization factor) between $l_2(V)$ and
$L_2(\TT^n,\CC^{|W|})$.

After this transform, $A$ becomes the operator of multiplication
in $L_2(\TT^n,\CC^{|W|})$ by a rational $|W|\times |W|$ matrix
function $A(z)$. This means that non-zero $l_2$-solutions of
$Au=0$ are in one-to-one correspondence with $\CC^{|W|}$-valued
$L_2$-functions $\hat{u}$ on $\TT^n$ such that $A(z)\hat{u}(z)=0$
a.e. on $\TT^n$. Since we assumed that $u$, and hence $\hat{u}$ is
not a zero element of $l_2$, we can conclude that the set of
points of the torus $\TT^n$ over which the matrix $A(z)$ has a
non-trivial kernel, has a positive measure. On the other hand,
this set in $\CC^n$ is given by the algebraic equation $\det
A(z)=0$ and thus is algebraic. The only way it can intersect the
torus over a subset of a positive measure is that it coincides
with the whole space $\CC^n$. Hence, $A(z)$ has a non-zero kernel
at any point $z$. Thus, its determinant is identically equal to
zero. Considering this matrix over the field $\mathcal{Q}$ of
rational functions, one can apply the standard linear algebra
statement that claims existence of a non-zero rational solution
$\phi (z)$ of $A(z)\phi (z)=0$. As indicated before, such
functions before the Floquet transform were compactly supported
solutions of $Au=0$. This proves the first statement of the
theorem, about the existence of compactly supported
eigenfunctions.

To prove completeness, we need to do a little bit more work. Let
us denote by $Q_1(z),...,Q_r(z)$ a finite set of the generators of
all non-zero polynomial (vector-valued) solutions of $A(z)Q(z)=0$
(it is known to exist, e.g. \cite[lemma 7.6.3, Ch.VII]{Horm}).
Floquet transform reduces the completeness statement we need to
prove to the following
\begin{lemma}\label{L:completeness}
Combinations
\begin{equation}\label{E:sections}
y(z)=\sum\limits_{j=1,..,r} a_j(z)Q_j(z),
\end{equation}
where $a_j(z)$ are finite Laurent sums, are $L_2$-dense in the
space of all $\CC^{|W|}$-valued $L_2$-solutions of the equation
\begin{equation}\label{E:Ay=0}
A(z)y(z)=0.
\end{equation}
\end{lemma}
{\em Proof of the lemma}. First of all, any $L_2(\TT^n)$-function
$a_j$ can be approximated by a finite Laurent sum. Indeed, this is
done by taking finite partial sums of the Fourier series of $a_j$
on the torus $\TT^n$. So, it is sufficient to approximate any
$L_2$-solution $y(z)$ of (\ref{E:Ay=0}) by sums (\ref{E:sections})
with $L_2$ coefficients $a_j$. Let $k>0$ be the minimal (over
$z\in\CC^n$ or $z\in\TT^n$, which is the same) dimension of
$KerA(z)$. The set $B\subset \TT^n$ of points $z$ where $dim\,
Ker\, A(z)>k$ is an algebraic variety of codimension at least $2$,
and hence has zero measure on $\TT^n$. Hence, it is sufficient to
do $L_2$ approximation outside of small neighborhoods of $B$. Let
$z_0\in\TT^n \backslash B$ and $U$ be a sufficiently small
neighborhood of $z_0$ not intersecting $B$. Then over (a complex
neighborhood of) $U$ the kernels $Ker A(z)$ form a trivial
holomorphic vector bundle. Let $f_l(z)$ be a basis of holomorphic
sections of this bundle. Then the portion of $y$ over $U$ can be
represented as $\sum b_l(z)f_l(z)$ with $L_2$-functions $b_l$.
Now, one uses \cite[lemma 7.6.3, Ch. VII]{Horm} again to see that
sums (\ref{E:sections}) with analytic $a_j$ approximate the
sections $f_l$. This proves the Lemma and hence the Theorem. \qed

The following observation is standard:
\begin{proposition}
If the periodic operator $A$ is self-adjoint, then its spectrum
has no singular continuous part.
\end{proposition}
Indeed, the singular continuous part is excluded for such periodic
operators by the standard well known argument (e.g.,
\cite{GN,Thomas}, or the proof of Theorem 4.5.9 in \cite{Ku93}).

Now the case of quantum graphs (at least when the Dirichlet
spectrum is excluded) can be reduced to the combinatorial one,
similarly to the way described in \cite{Ku04_graphs1}.

\begin{theorem}\label{bound_quantum}
Let $\GG$ be a $G=\ZZ^n$-periodic (in the meaning already
specified) quantum graph equipped with the second derivative
Hamiltonian and arbitrary vertex conditions at the vertices. Then,
existence of a non-zero $L_2$-eigenfunction corresponding to an
eigenvalue $\lambda$ implies existence of a compactly supported
eigenfunction, and the set of compactly supported eigenfunctions
is complete in the eigenspace. If the vertex conditions are
self-adjoint, the spectrum of the Hamiltonian has no singular
continuous part.
\end{theorem}
\proof The first step is to make sure that $\lambda$ stays away
from the Dirichlet spectrum $\sigma_D$, which in the case we
consider is discrete. If by any chance $\lambda \in \sigma_D$, one
can introduce ``fake'' additional vertices of degree $2$ on the
edges of the fundamental domain of the graph and then repeat them
periodically in such a way that the Dirichlet eigenvalues of the
new shorter edges will avoid $\lambda$. If one imposes Neumann
conditions at these new vertices, their introduction does not
influence the operator at all. So, we can assume from the start
that $\lambda$ is not in $\sigma_D$. Let now $F$ be an
$L_2$-eigenfunction. Since we are away from the Dirichlet spectrum
$\sigma_D$, resolvent and trace estimates analogous to the ones in
the proof of the previous theorem show that the vector
$f=\{F(v)\}$ of the vertex values belongs to $l_2(V)$ if and only
if $F\in\L_2(\GG)$. Since $\lambda$ is not in $\sigma_D$, solving
the boundary value problem for the eigenfunction equation
$HF=\lambda F$ on each edge separately in terms of the boundary
values of $F$, we can express the derivatives of $F$ at each
vertex in terms of its vertex values $f$ solely. Thus, boundary
conditions (which involve the values of $F$ and of its vertex
derivatives) lead to a periodic finite order difference equation
$Af=0$ on the combinatorial counterpart of the quantum graph.
Theorem \ref{T:bound_combinat} claims existence and completeness
of combinatorial compactly supported solutions. Reversing the
procedure (which is possible since we are not on the Dirichlet
spectrum), we conclude existence and completeness of compactly
supported eigenfunctions of the quantum graph.

The part about the absence of singular continuous spectrum is
standard (as for the combinatorial graphs). \qed
\begin{remark}\label{R:gaps}Compactly supported eigenfunctions on graphs are sometimes
  called ``scars.''
\end{remark}

\section{Acknowledgment\label{S:Acknow}}

The author thanks M.~Aizenman, R.~Carlson, P.~Exner,
R.~Grigorchuk, S.~Novikov, H.~Schenck, J.~Schenker, and
R.~Schrader for relevant information and M.~Solomyak and the
reviewers for useful comments about the manuscript.

This research was partly sponsored by the NSF through the Grants
DMS 9610444, 0072248, 0296150, and 0406022. The author expresses
his gratitude to NSF for this support. The content of this paper
does not necessarily reflect the position or the policy of the
federal government, and no official endorsement should be
inferred.

\end{document}